# Mapping metallic *d*-wave altermagnetism across the TiNiSi structural family

Zhen Zhang[1,*], Subhadip Pradhan[2], Kirill D. Belashchenko[2], Vladimir Antropov[1,3,*]

[1]*Department of Physics and Astronomy, Iowa State University, Ames, IA 50011, USA*

[2]*Department of Physics and Astronomy and Nebraska Center for Materials and Nanoscience, University of Nebraska-Lincoln, Lincoln, Nebraska 68588, USA*

[3]*Ames National Laboratory, U.S. Department of Energy, Ames, IA 50011, USA*

[*]Corresponding authors: Zhen Zhang zhenz1@iastate.edu, Vladimir Antropov antropov@iastate.edu

**Abstract**

The prospect of using altermagnets as switchable sources of perpendicularly polarized spin currents has intensified the search for candidate materials, yet metallic *d*-wave systems with sizable spin-splitter responses remain scarce. Here, we identify an empirical magnetic motif that supports metallic *d*-wave altermagnetism in the TiNiSi structural family: ferromagnetically ordered zigzag chains with antiferromagnetic interchain coupling in a relatively low-symmetry crystal environment. The TiNiSi structure type combines this motif with broad chemical flexibility and competing magnetic ground states. By combining first-principles screening of thermodynamic stability and magnetic ground states across 280 ternary systems, we identify 16 metallic *d*-wave altermagnets. This set recovers four experimentally known members—WFeB, NbMnP, TaMnP, and NbMnAs—and yields 12 new predictions, of which ScMnP, TaMnAs, ScMnAs, MoMnAs, MoMnSi, and WMnSi are the most promising. Noncollinear calculations indicate that the collinear altermagnetic configuration is the ground state across them. Each of the six leading new candidates has a nonzero spin-splitter angle and a finite anomalous Hall conductivity. Notably, ScMnP and TaMnAs exhibit strong spin-splitter responses despite modest altermagnetic band splitting. These results establish TiNiSi-type metallic *d*-wave altermagnets as a chemically versatile platform for efficient charge-to-spin conversion and provide an empirical magnetic-motif-guided route to identifying further candidates.

## Introduction

Altermagnetism has emerged as a third class of collinear magnetic order, distinct from ferromagnetism and antiferromagnetism. It combines nonrelativistic spin-split electronic bands, as in a ferromagnet, with the vanishing net magnetization of an antiferromagnet[1–4]. This combination makes altermagnets attractive for generating and detecting spin currents without large net magnetization. Without spin-orbit coupling (SOC) or strain, however, the altermagnetic (AM) symmetry of *g*-wave[1] systems, including MnTe[5] and CrSb[6], forces the spin-conductivity tensor to cancel, so the current is unpolarized in every direction. By comparison, *d*-wave[1] symmetry permits a nonzero spin-splitter response—a pure transverse spin current—in the metallic phase[7–10]. Such a response offers a route to switching perpendicular magnetization in spin-orbit-torque-like devices, an important function for dense magnetic memory[11]. Although first proposed as a *d*-wave altermagnet, $RuO_2$ is now considered nonmagnetic in bulk[12–14]. Experimental evidence for metallic *d*-wave altermagnetism has so far been found in only a few systems, notably $Mn_5Si_3$[15–17] and tetragonal oxychalcogenides[18,19]. These few experimentally verified examples underscore the scarcity of experimentally established metallic *d*-wave altermagnets.

In our recent work, combined experimental and theoretical evidence identified WFeB as a metallic *d*-wave altermagnet within the broad TiNiSi-type structural family, which offers a symmetry framework for realizing spintronic functionality based on AM metals with relatively high Néel temperatures[20]. That study also showed that the family's spin-group symmetry can produce a large nonrelativistic spin-splitter response even when the band spin splitting is only about 100 meV. The same symmetry permits deterministic electrical reorientation of the magnetic domain through current-induced staggered torques in a thin film. Together, these ingredients allow an electrically reconfigurable source of spin current polarized normal to the film, offering an alternative to conventional spin-Hall injectors in spin-torque devices. Realizing this opportunity requires a broader selection of metallic *d*-wave altermagnets with appreciable spin-splitter responses.

Fortunately, altermagnets are well suited for high-throughput discovery, either using density functional theory (DFT)[21–27] or artificial intelligence (AI) assisted approaches[28–31], based on three computationally accessible characteristics: crystal symmetry, collinear compensated magnetic order, and nonrelativistic spin-split band structure. These characteristics can be enumerated, and their computations can be automated, enabling high-throughput DFT or AI-assisted screening. Incorporating symmetry constraints into high-throughput workflows can substantially narrow the candidate space entering costly first-principles calculations. High-throughput DFT is particularly efficient and effective when applied to a predefined structural family. We previously used high-throughput DFT to identify magnet and superconductor families, such as *g*-wave boride altermagnets[21], dimerized quantum magnets[32,33], kagome magnets[34,35], and ambient-pressure hydride superconductors[36]. In particular, we previously demonstrated that once a crystal structure and its collinear magnetic ground state meet the symmetry requirements for altermagnetism, altermagnets can emerge as a family crystallizing in the same structural type[21].

Motivated by the need of seeking metallic *d*-wave altermagnets, we perform high-throughput first-principles screening of thermodynamic stability and magnetic ground states in TiNiSi-type MTA compounds (M = Sc, Ti, Y, Zr, Nb, Mo, La, Hf, Ta, W; T = Cr, Mn, Fe, Co; A = N, P, As, C, Si, Ge, B). The search yields 16 metallic *d*-wave altermagnets: it recovers the four known members (WFeB, NbMnP, TaMnP, and NbMnAs) and predicts 12 additional candidates. We further evaluate the spin-splitter response and anomalous Hall effect (AHE) for six representative new *d*-wave altermagnets—ScMnP, TaMnAs, MoMnAs, ScMnAs, MoMnSi, and WMnSi.

## Results

### Structure and phase stability

Compounds adopting the TiNiSi structure (orthorhombic *Pnma* space group; also called the SrMgSi type) occur widely among group-15 (N, P, As, Sb, Bi), group-14 (C, Si, Ge, Sn), and group-13 (B, Al, Ga, In) materials[37]. We therefore began with equiatomic MTA compounds covering A = N, P, As, Sb, Bi, C, Si, Ge, Sn, B, Al, Ga, and In. Since the atomic radii $R_{\mathrm{Ti}} > R_{\mathrm{Ni}} > R_{\mathrm{Si}}$ , compounds stabilized in this structural type generally obey $R_{\text{element on the Ti site}} > R_{\text{element on the Ni site}} > R_{\text{element on the Si site}}$ . In particular, known altermagnets and related antiferromagnets in this family, including WFeB[20], NbMnP[38,39], TaMnP[40], and NbMnAs[41], all share the following arrangement: the magnetic element occupies the Ni site, and the atomic radii obey $R_{\text{element on the Ti site}} > R_{\text{element on the Ni site}} > R_{\text{element on the Si site}}$. In addition, Mn and Fe warrant particular consideration because they often support robust magnetic moments and may give rise to relatively high Néel temperatures.

On this basis, the subsequent high-throughput search considered TiNiSi-type MTA phases with magnetic T element and $R_{\text{M on the Ti site}} > R_{\text{T on the Ni site}} > R_{\text{A on the Si site}}$. We selected M = Sc, Ti, Y, Zr, Nb, Mo, La, Hf, Ta, and W, and T from the 3d magnetic elements Cr, Mn, Fe, and Co. Because Sb, Bi, Sn, Al, Ga, and In are larger than the 3d magnetic elements, we excluded them and retained A = N, P, As, C, Si, Ge, and B. Elementally substituted MTA phases that violate $R_{\text{M on the Ti site}} > R_{\text{T on the Ni site}} > R_{\text{A on the Si site}}$ are likely to be unstable and difficult to synthesize in the intended TiNiSi structure. Applying this filter substantially reduced the number of MTA phases passed to the first-principles screening.

Polymorphism must also be considered when predicting equiatomic ternaries. Seven structure types—TiNiSi, ZrAlNi, PbFCl, LiGaGe, YPtAs, UGeTe, and LaPtSi—account for 91% of the known equiatomic ternaries[42]. The two most prevalent, TiNiSi and ZrAlNi (hexagonal $P\bar{6}2m$ space group), alone represent 69% of known equiatomic ternaries[42]. An exhaustive treatment of every possible polymorph is not feasible in a high-throughput first-principles study. We therefore included the two dominant structure types in the stability calculations to obtain a more reliable comparison of relative energies and convex-hull distances. Tests of the other five common structure types in representative arsenide and boride systems found none to be more stable than these two choices. Because $R_{\mathrm{Zr}} > R_{\mathrm{Al}} > R_{\mathrm{Ni}}$ and our screening already imposes $R_{\mathrm{M}} >$ $R_{\mathrm{T}} > R_{\mathrm{A}}$, substitutions into the ZrAlNi structure also follow $R_{\mathrm{M\ on\ the\ Zr\ site}} > R_{\mathrm{T\ on\ the\ Al\ site}} > R_{\mathrm{A\ on\ the\ Ni\ site}}$, keeping the constructed phases chemically reasonable in terms of atomic size.

Figures 1(a) and 1(b) depict the TiNiSi- and ZrAlNi-type structures, respectively. To highlight their differences, we show the metal–metal bonds in the *ac* plane of the TiNiSi structure and the *ab* plane of the ZrAlNi structure. In both structures, the metal atoms form trigonal prisms. The larger M atoms occupy corner-sharing sites, whereas the smaller T atoms sit at the non-corner-sharing sites of the prisms[42,43].

The thermodynamic stability of the constructed TiNiSi- and ZrAlNi-type phases was assessed from ternary convex-hull calculations. Reference phases on each hull were drawn from the Materials Project[44] database, and their formation energies were recalculated using spin-polarized DFT. For each TiNiSi-type compound, we used the energy of the magnetic ground state identified below; for its ZrAlNi-type counterpart, we used the FM-state energy. The latter structure contains equilateral triangles of magnetic T atoms and may therefore support frustration and noncollinear order. Tests on ZrAlNi-type TaMnAs, TiMnAs, ZrMnAs, and HfMnAs nevertheless showed that noncollinearity lowers the ZrAlNi-type energy by much less than the energy separation between the two structure types and thus does not change their predicted ordering. Because this study targets the altermagnetically relevant TiNiSi structure and full noncollinear calculations are impractical at high-throughput scale, the FM approximation is adequate for evaluating the competing ZrAlNi phase. The resulting ZrAlNi stability data may also inform future studies of magnetism in that structure.

The distances above the reconstructed convex hull, $E_d$, for the TiNiSi- and ZrAlNi-type phases are shown as blue and red poles, respectively, in Figs. 1 and 2 for A = P, As, Si, Ge, and B. A stable phase on the hull has $E_d = 0$, whereas $E_d \leq 0.2$ eV/atom is commonly used as an upper bound for potentially accessible metastable compounds[45]. We also applied a tighter $E_d \leq 0.05$ eV/atom cutoff to identify the candidates with the strongest synthesis prospects. This procedure yields many stable or metastable compounds for A = P, As, Si, Ge, and B. By contrast, most nitrides and carbides in either structure have large $E_d$ values (see Fig. S1 of the Supplementary Information). The likely reason is that N and C are too small to support these frameworks, with B being the smallest element that commonly forms them. We therefore excluded nitrides and carbides from the subsequent magnetic-ground-state calculations.

For A = P, As, Si, and Ge, the TiNiSi structure is generally more stable than the ZrAlNi alternative, as indicated by the larger number of absent or shorter blue poles relative to their red counterparts (Figs. 1 and 2). Among borides, the numbers of compositions favoring the two structures are comparable. For example, both calculation and experiment place WFeB, MoCoB, and WCoB in the TiNiSi structure[20,46], whereas NbFeB and TaFeB are predicted and observed in the ZrAlNi structure[47]. Overall, the predictions agree well with available syntheses. For the altermagnetically relevant TiNiSi structure, we identify 30 (36), 19 (30), 30 (40), 26 (40), and 16 (37) compounds with $0 \leq E_d \leq 0.05$ eV/atom ($0 \leq E_d \leq 0.2$ eV/atom) for A = P, As, Si, Ge, and B, respectively. Collectively, these stable and metastable compositions define a broad TiNiSi-type materials family.

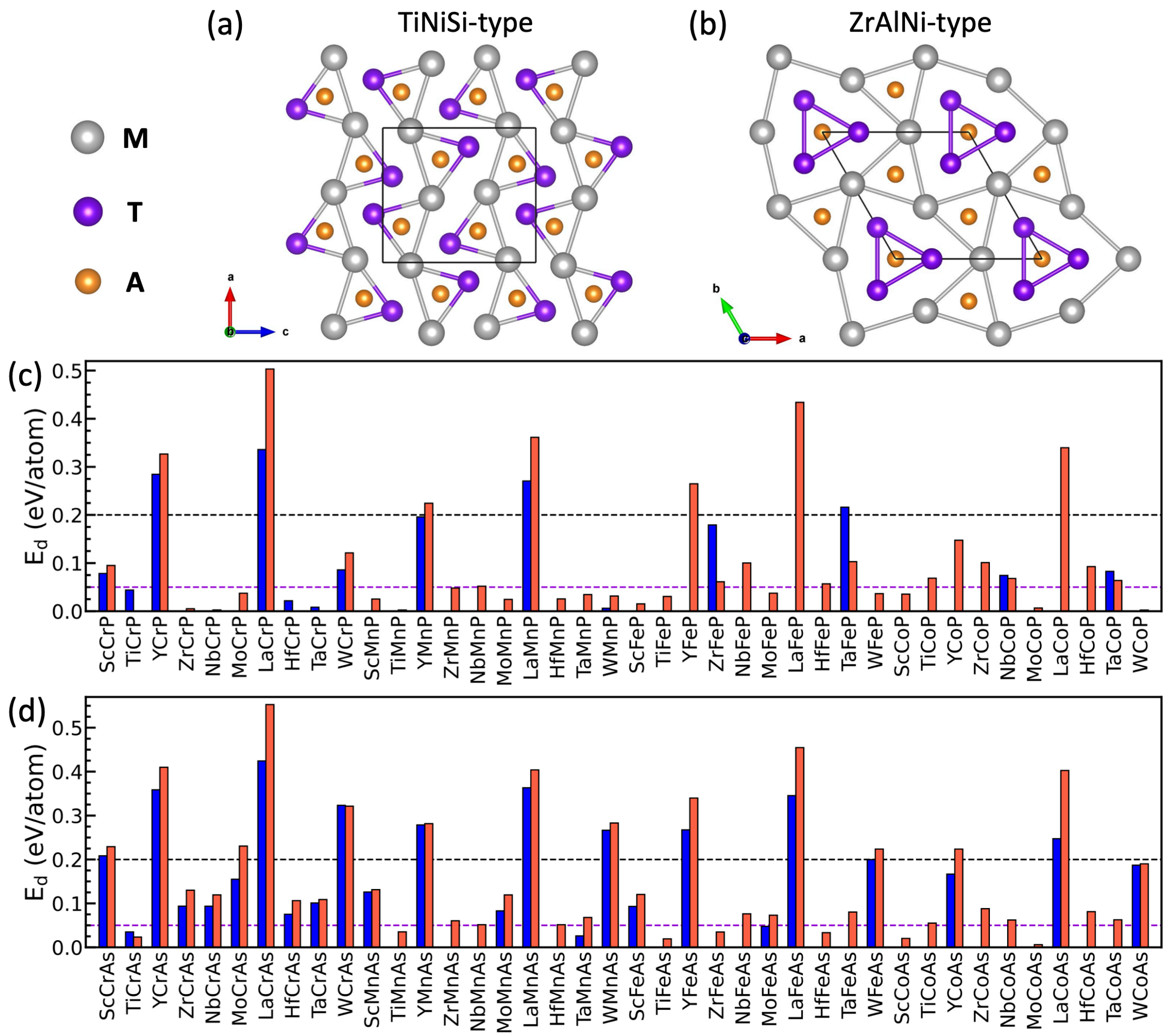

**Fig. 1. Crystal structures and convex-hull distances for MTA compounds with A = P and As.** Unit cells of the (a) TiNiSi- and (b) ZrAlNi-type structures. Hull distances for (c) phosphides and (d) arsenides. Blue and red poles denote TiNiSi- and ZrAlNi-type phases, respectively. The criteria $E_d \leq 0.2$ eV/atom and $E_d \leq 0.05$ eV/atom are marked by the black and purple horizontal dashed lines, respectively.

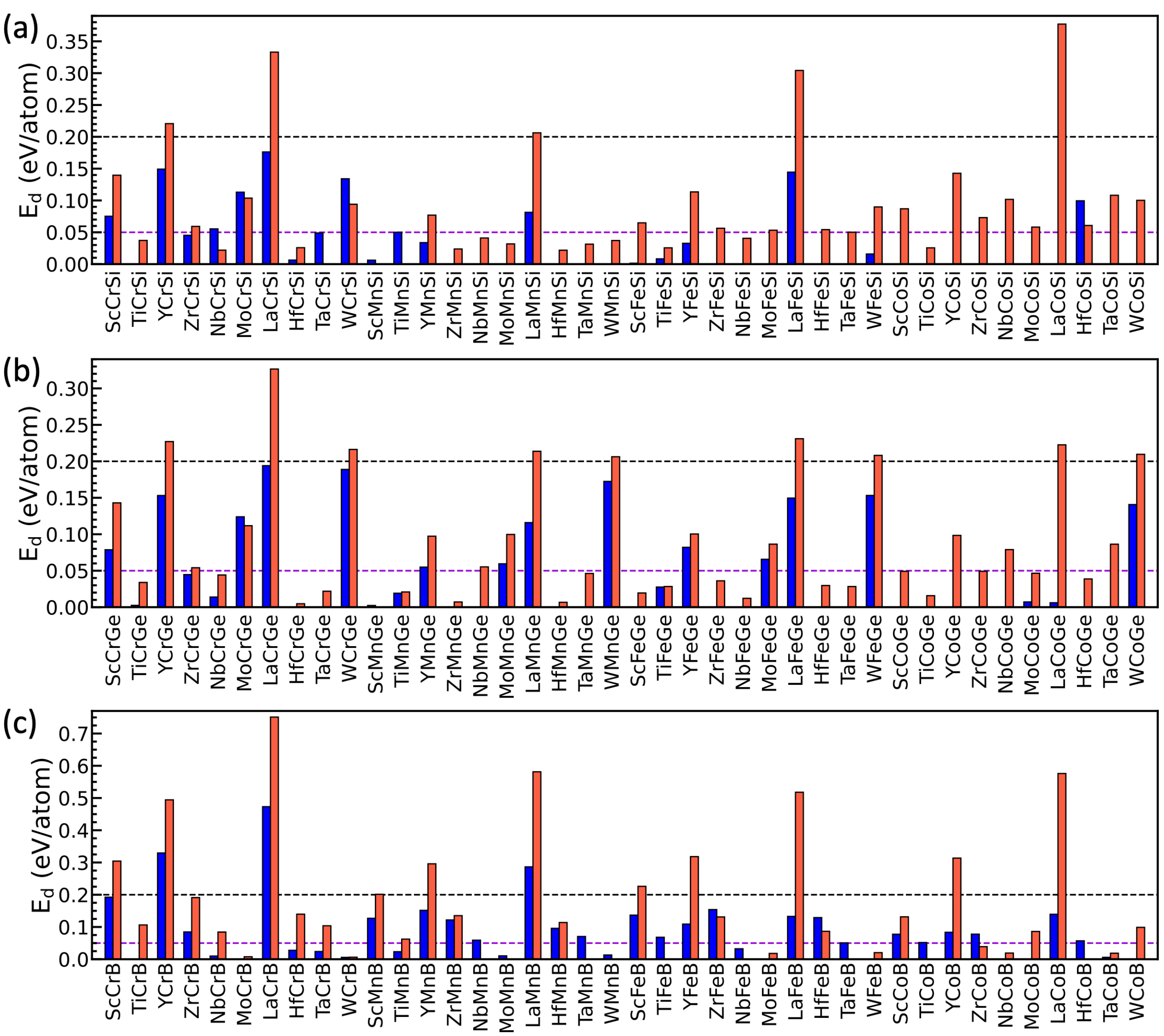

**Fig. 2. Convex-hull distances for MTA compounds with A = Si, Ge, and B.** Hull distances for (a) silicides, (b) germanides, and (c) borides. Blue and red poles denote TiNiSi- and ZrAlNi-type phases, respectively. The criteria $E_d \leq 0.2$ eV/atom and $E_d \leq 0.05$ eV/atom are marked by the black and purple horizontal dashed lines, respectively.

We emphasize that the reported $E_d$ values come from convex hulls built from the constructed TiNiSi- and ZrAlNi-type phases together with known convex-hull phases in the Materials Project[44]. We selected these two candidate structures because the literature identifies them as the most common equiatomic ternary structure types. This choice keeps the screening closely connected to experimentally relevant synthesis targets. Nonetheless, some studied compositions may adopt other known or as-yet-unreported structures with comparable or lower energies. A global structure search for every composition lies outside the scope of this work. Accordingly, $E_d$ should be interpreted as a measure of the likelihood of obtaining a given composition in the TiNiSi structure, not as a binary prediction of synthesizability. Polymorphism is common among equiatomic ternaries when competing structures are close in energy[42], and different synthesis routes or conditions can stabilize different structures at the same composition.

**Magnetic ground state**

For each of the 183 TiNiSi-type phases with $E_d \leq 0.2$ eV/atom, we compared the total energies of several collinear magnetic configurations to determine the magnetic ground state. The four magnetic configurations within the unit cell—ferromagnetic (FM), altermagnetic (AM), antiferromagnetic-1 (AFM-1), and AFM-2—are illustrated in Fig. 3(a). To host *d*-wave altermagnetism, a TiNiSi-type compound must exhibit collinear magnetic order belonging to the nontrivial ${}^2m{}^1m{}^2m$ spin point group. The nearest-neighbor magnetic atoms form zigzag chains extending along the *b* axis, and the AM configuration compatible with this symmetry requirement consists of ferromagnetically aligned spins within each zigzag chain and AFM coupling between adjacent chains, as further illustrated in Fig. S2 of the Supplementary Information. In contrast, both conventional AFM-1 and AFM-2 configurations feature AFM alignment within the zigzag chains.

Including the nonmagnetic (i.e., non-spin-polarized) configuration, we relaxed and evaluated five states for every compound. Figures 3 and 4 show the magnetic solutions, i.e., converged magnetic configurations with well-defined magnetic moments, for A = P, As, Si, Ge, and B, respectively. The upper panels give the energy of each magnetic solution relative to the nonmagnetic (NM) state. Because every magnetic configuration was relaxed independently, these differences contain both magnetic and structural contributions. The lower panels show the magnetic moment on the 3d transition-metal atom T. To display thermodynamic stability, compounds with $0 \leq E_d \leq 0.05$ eV/atom and $0.05 < E_d \leq 0.2$ eV/atom are marked by purple and black composition labels, respectively, in Figs. 3 and 4. If all attempted magnetic configurations for a compound converge to the nonmagnetic state and no magnetic solution is obtained, the compound is classified as nonmagnetic. Altogether, the screening yields 99 magnetic and 84 nonmagnetic compounds. The lowest-energy solution defines the magnetic ground state of each magnetic compound. We divide the magnetic compounds into three groups—*d*-wave altermagnets, ferromagnets, and conventional antiferromagnets—using the colors in Figs. 3 and 4. The nonmagnetic compounds are provided in Table 1.

Because the generalized-gradient approximation (GGA) tends to overestimate magnetic moments, we recalculated the ground-state moments of every magnetic compound using the local density approximation (LDA). In 14 compounds—MoMnP, ZrFeP, ScCrP, TiFeAs, ZrFeAs, MoFeAs, HfFeAs, MoCrAs, NbCrSi, YFeSi, LaFeSi, TaFeGe, HfCrGe, and ScCrB—the moment collapses to zero in LDA. The nonmagnetic and the weakly magnetic compounds in DFT are expected to appear paramagnetic in experimental measurements. Nevertheless, these 14 weakly magnetic compounds may exhibit stronger spin fluctuations. Six of them—MoMnP, TiFeAs, ZrFeAs, MoFeAs, HfFeAs, and TaFeGe—are particularly interesting candidates for AM spin fluctuations and possibly AM-spin-fluctuations-mediated superconductivity[48].

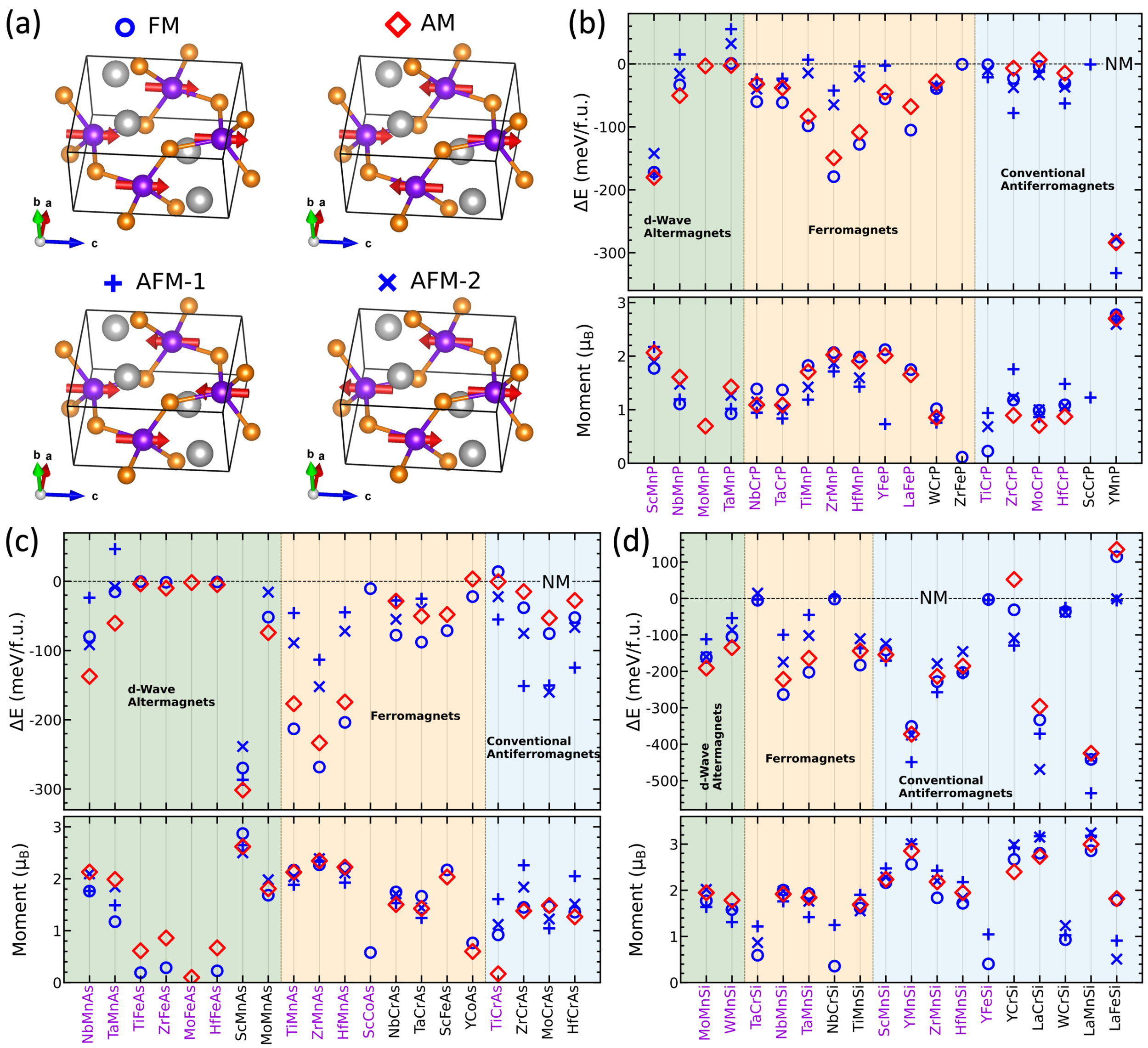

**Fig. 3. Magnetic configurations and magnetic solutions for TiNiSi-type MTA compounds with A = P, As, and Si within 0.2 eV/atom of the convex hull.** (a) Magnetic configurations with their symbols and labels. Total energies relative to the nonmagnetic solution (upper panel) and magnetic moments on T (lower panel) for (b) phosphides, (c) arsenides, and (d) silicides obtained by GGA. Purple and black composition labels denote compounds within 0.05 and 0.2 eV/atom of the convex hull, respectively.

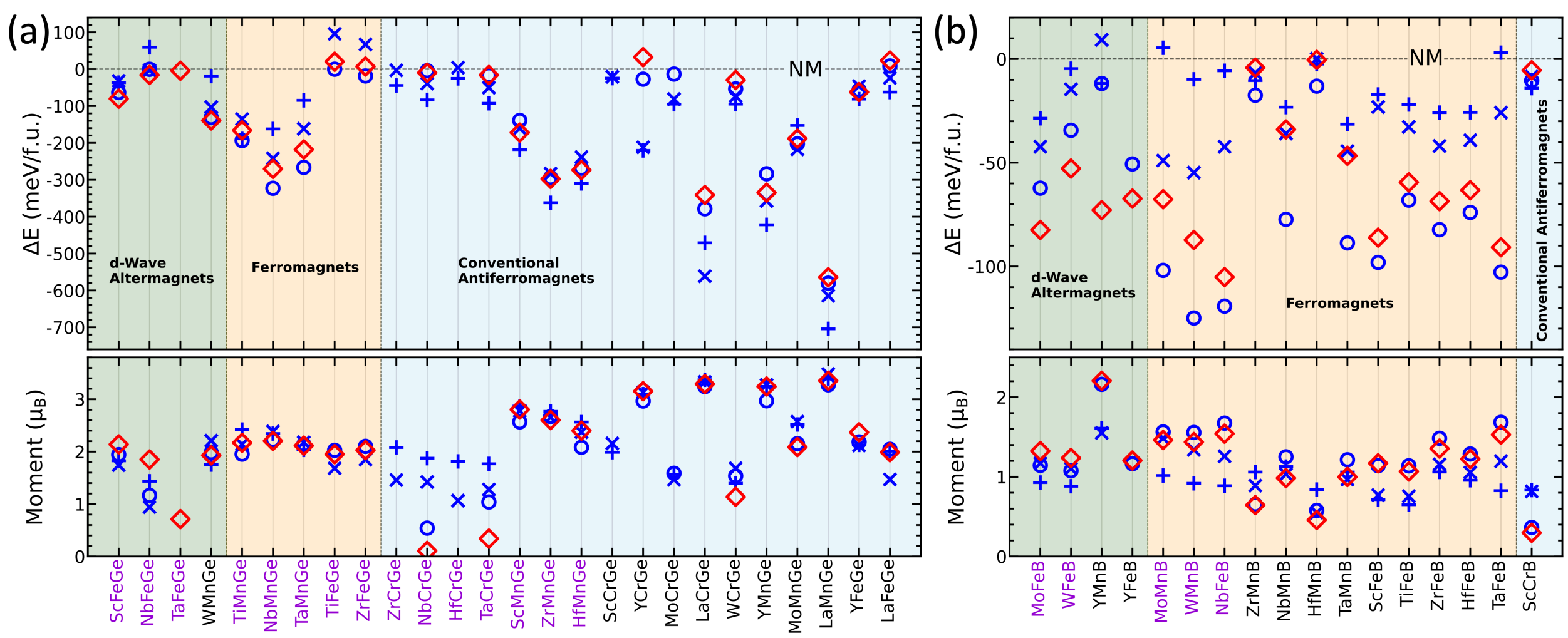

**Fig. 4. Magnetic solutions for TiNiSi-type MTA compounds with A = Ge and B within 0.2 eV/atom of the convex hull.** Total energies relative to the nonmagnetic solution (upper panel) and magnetic moments on T (lower panel) for (a) germanides and (b) borides obtained by GGA. Purple and black composition labels denote compounds within 0.05 and 0.2 eV/atom of the convex hull, respectively.

**Table 1. Nonmagnetic TiNiSi-type MTA compounds within 0.2 eV/atom of the convex hull.**

| Chemical families | $0 \leq E_d \leq 0.05$ eV/atom | $0.05 < E_d \leq 0.2$ eV/atom |
|---|---|---|
| phosphides | WMnP, ScFeP, TiFeP, NbFeP, MoFeP, HfFeP, WFeP, ScCoP, TiCoP, YCoP, ZrCoP, MoCoP, LaCoP, HfCoP, WCoP | NbCoP, TaCoP |
| arsenides | NbFeAs, TaFeAs, TiCoAs, ZrCoAs, NbCoAs, MoCoAs, HfCoAs, TaCoAs | WFeAs, WCoAs |
| silicides | TiCrSi, ZrCrSi, HfCrSi, ScFeSi, TiFeSi, ZrFeSi, NbFeSi, MoFeSi, HfFeSi, TaFeSi, WFeSi, ScCoSi, TiCoSi, YCoSi, ZrCoSi, NbCoSi, MoCoSi, LaCoSi, TaCoSi, WCoSi | ScCrSi, MoCrSi, HfCoSi |
| germanides | TiCrGe, HfFeGe, ScCoGe, TiCoGe, YCoGe, ZrCoGe, NbCoGe, MoCoGe, LaCoGe, HfCoGe, TaCoGe | MoFeGe, WFeGe, WCoGe |
| borides | TiCrB, NbCrB, MoCrB, HfCrB, TaCrB, WCrB, TiMnB, NbCoB, MoCoB, TaCoB, WCoB | ZrCrB, ScMnB, LaFeB, ScCoB, TiCoB, YCoB, ZrCoB, LaCoB, HfCoB |

**Table 2. The full list of the predicted AM TiNiSi-type compounds.** The predicted $E_d$ (eV/atom) of the TiNiSi-type structure and magnetic moment on transition metal atom T are shown in the second and third columns, respectively. The synthesized structure, experimental Néel temperature, neutron diffraction confirmation of the predicted magnetic structure, and AHE measurement available from literature are shown in the fourth, fifth, sixth, and seventh columns, respectively.

| **Predicted altermagnetic TiNiSi-type compounds** | **Predicted $E_d$ (eV/atom) of the TiNiSi-type structure** | **Predicted magnetic moment on T ($\mu_B$) by GGA (LDA)** | **Synthesized structure** | **Experimental Néel temperature (K)** | **Neutron diffraction confirmation of the predicted magnetic structure** | **Anomalous Hall effect measured** |
|---|---|---|---|---|---|---|
| ScMnP | 0 | 2.06 (1.58) | – | – | – | – |
| NbMnP | 0 | 1.61 (1.25) | TiNiSi-type[38,39] | 233[38,39] | [38] (strongly noncollinear) | [39] |
| TaMnP | 0 | 1.43 (1.07) | TiNiSi-type[40] | 220–240[40] | [40] | [40] |
| NbMnAs | 0 | 2.13 (1.77) | TiNiSi-type[41] | 354[41] | – | [41] |
| TaMnAs | 0.025 | 1.99 (1.63) | – | – | – | – |
| ScMnAs | 0.125 | 2.62 (2.24) | – | – | – | – |
| MoMnAs | 0.082 | 1.81 (1.37) | – | – | – | – |
| MoMnSi | 0 | 1.95 (1.63) | – | – | – | – |
| WMnSi | 0 | 1.79 (1.49) | – | – | – | – |
| ScFeGe | 0 | 2.14 (1.85) | ZrAlNi-type[49,50] | – | – | – |
| NbFeGe | 0 | 1.85 (1.48) | TiFeSi-type[51] | – | – | – |
| WMnGe | 0.172 | 1.93 (1.60) | – | – | – | – |
| MoFeB | 0 | 1.32 (1.05) | – | – | – | – |
| WFeB | 0 | 1.23 (0.99) | TiNiSi-type[20] | 240[20] | [20] | – |
| YMnB | 0.151 | 2.21 (1.71) | – | – | – | – |
| YFeB | 0.109 | 1.21 (0.93) | – | – | – | – |

The remaining 85 compounds retain well-defined magnetic moments. Sixteen are $d$-wave altermagnets: ScMnP, NbMnP, TaMnP, NbMnAs, TaMnAs, ScMnAs, MoMnAs, MoMnSi, WMnSi, ScFeGe, NbFeGe, WMnGe, MoFeB, WFeB, YMnB, and YFeB. Table 2 compares their predicted properties with the available experimental results. Within the assumption of collinear order, the screening recovers the four experimentally established members of this family—WFeB, NbMnP, TaMnP, and NbMnAs—providing an internal validation. The next section addresses their possible noncollinearity. The predicted hull stability ($E_d = 0$) agrees with the reported synthesis of WFeB[20], NbMnP[38,39], TaMnP[40], and NbMnAs[41] in the TiNiSi structure. Neutron diffraction supports the calculated AM ground state for WFeB[20] and TaMnP[40]. Neutron diffraction for NbMnP[38] indicates a strongly noncollinear ground state, which can be expressed as a combination of AM $B_{3g}$ irreducible representation and AFM-2 $B_{2u}$ irreducible representation.

Beyond the four experimentally known members, we predict 12 altermagnets: ScMnP, TaMnAs, ScMnAs, MoMnAs, MoMnSi, WMnSi, ScFeGe, NbFeGe, WMnGe, MoFeB, YMnB, and YFeB. Older studies report ScFeGe[49,50] and NbFeGe[51] in the ZrAlNi- and TiFeSi-type structures, respectively, even though $E_d = 0$ is predicted for their TiNiSi polymorphs. Because those reports date from three to five decades ago and have not been revisited recently, new synthesis would be valuable for checking their structures and possible polymorphism. MoFeB is likewise predicted to be stable, but our earlier experimental exploration of the Mo–Fe–B system did not detect this phase[21]. Its synthesis may therefore be difficult. Challenges are also likely for the refractory-boron systems YMnB and YFeB and for WMnGe, which has a relatively large $E_d$ of 0.172 eV/atom. These considerations leave six leading candidates: ScMnP, TaMnAs, ScMnAs, MoMnAs, MoMnSi, and WMnSi.

These altermagnets carry robust magnetic moments of roughly 1–2 $\mu_B$ on the 3d transition-metal atom T, with LDA values slightly below the GGA results (Table 2). In addition to the ground-state local moment, the variation in local-moment magnitude among the magnetic solutions provides a qualitative indication of moment localization and, consequently, of the suitability of a fixed-length Heisenberg description. As shown in Figs. 3 and 4, WFeB, NbMnP, TaMnP, NbMnAs, ScMnP, ScMnAs, MoMnAs, MoMnSi, and WMnSi exhibit relatively small variations in local-moment magnitude, consistent with more localized magnetism, whereas TaMnAs shows a stronger configuration dependence, suggesting a more itinerant character. This interpretation should nevertheless be treated with caution because each magnetic configuration was structurally relaxed independently; the observed variations therefore also contain contributions from magnetostructural coupling.

For the four known compounds, the GGA energy separations between the AM ground state and the next-lowest state are 18, 17, 2, and 46 meV/f.u. for WFeB, NbMnP, TaMnP, and NbMnAs, respectively. Their experimental Néel temperatures ($T_N$) are 240[20], 233[38,39], 220–240[40], and 354 K[41], respectively. For the six newly predicted compounds, the corresponding energy separations are 5, 45, 15, 22, 29, and 30 meV/f.u. for ScMnP, TaMnAs, ScMnAs, MoMnAs, MoMnSi, and WMnSi, respectively. These energy scales suggest robust altermagnetism and $T_N$ values that may approach or exceed room temperature. Note that the energy differences among different magnetic solutions presented in this work were obtained by GGA, which is generally good at predicting the collinear magnetic ground state and the geometry. The energy differences reported previously for WFeB were obtained by LDA because it yielded a magnetic moment in closer agreement with the experimental value[20].

To estimate $T_N$ quantitatively, we calculated magnetic exchange couplings and used a mean-field approximation[20,21,32,35]. In our earlier WFeB study, retaining the first four couplings in LDA—two intrachain FM and two interchain AFM interactions—yielded $T_N = 230$ K, close to the experimental value of 240 K for WFeB[20]. Applying the same approach gives $T_N$ values of 270, 200, and 420 K for NbMnP, TaMnP, and NbMnAs, respectively, in reasonable agreement with the experimental values above. For the six new candidates, the estimated $T_N$ values are 530, 400, 610, 230, 460, and 450 K for ScMnP, TaMnAs, ScMnAs, MoMnAs, MoMnSi, and WMnSi, respectively. Mean-field predictions, however, can be >20% higher than corresponding Monte Carlo estimates.

**Table 3. The list of the predicted FM TiNiSi-type compounds within 0.05 eV/atom of the convex hull.** The predicted $E_d$ (eV/atom) of the TiNiSi-type structure and magnetic moment on transition metal atom T are shown in the second and third columns, respectively. The synthesized structure and experimental confirmation of the predicted ferromagnetism in the TiNiSi-type structure are shown in the fourth and fifth columns, respectively.

| Predicted ferromagnetic TiNiSi-type compounds | Predicted $E_d$ (eV/atom) of the TiNiSi-type structure | Predicted magnetic moment on T ($\mu_B$) by GGA (LDA) | Synthesized structure | Confirmation of the predicted ferromagnetism in the TiNiSi-type structure |
|---|---|---|---|---|
| NbCrP | 0 | 1.39 (1.03) | TiNiSi-type[52] | NM[52] |
| TaCrP | 0.008 | 1.37 (1.02) | TiNiSi-type, ZrAlNi-type, and *P12₁/c1*[53] | – |
| TiMnP | 0 | 1.82 (1.57) | ZrAlNi-type[54] | – |
| ZrMnP | 0 | 2.07 (1.84) | TiNiSi-type[55] | FM[55] |
| HfMnP | 0 | 1.98 (1.76) | TiNiSi-type[55] | FM[55] |
| YFeP | 0 | 2.12 (1.93) | – | – |
| LaFeP | 0 | 1.75 (1.49) | – | – |
| TiMnAs | 0 | 2.17 (1.94) | ZrAlNi-type[56] | – |
| ZrMnAs | 0 | 2.27 (2.06) | – | – |
| HfMnAs | 0 | 2.20 (2.01) | – | – |
| ScCoAs | 0 | 0.58 (0.39) | – | – |
| TaCrSi | 0.048 | 0.59 (0.20) | – | – |
| NbMnSi | 0 | 2.00 (1.78) | ZrAlNi-type[57] | – |
| TaMnSi | 0 | 1.93 (1.75) | ZrAlNi-type[58] and TiFeSi-type[51] | – |
| TiMnGe | 0.019 | 1.95 (1.64) | – | – |
| NbMnGe | 0 | 2.23 (2.01) | ZrAlNi-type[59] | – |
| TaMnGe | 0 | 2.14 (1.93) | ZrAlNi-type[58] and TiFeSi-type[51] | – |
| TiFeGe | 0.027 | 2.03 (1.83) | TiFeSi-type[60] | – |
| ZrFeGe | 0 | 2.10 (1.94) | TiNiSi-type[58] | – |
| MoMnB | 0.010 | 1.56 (1.25) | – | – |
| WMnB | 0.013 | 1.56 (1.32) | – | – |
| NbFeB | 0.032 | 1.67 (1.41) | ZrAlNi-type[47] | – |

We also identify 22 (37) ferromagnets and 15 (32) conventional antiferromagnets with well-defined magnetic moments and $0 \leq E_d \leq 0.05$ eV/atom ($0 \leq E_d \leq 0.2$ eV/atom) in the TiNiSi family. Most of them show substantial magnetic stability: the ground state lies tens of meV/f.u. below the next-lowest state, and the local moment is about 1–3 $\mu_B$. The 22 predicted TiNiSi-type ferromagnets within 0.05 eV/atom of the convex hull are listed in Table 3. Among them, ZrMnP and HfMnP have been synthesized in the predicted TiNiSi-type structure, and experimental measurements have confirmed their predicted FM ground states, with Curie temperatures of 370 and 320 K, respectively; both compounds also exhibit decent magnetic anisotropy[55]. ZrFeGe has also been synthesized in the predicted TiNiSi-type structure[58], but its magnetic properties have not yet been measured. NbCrP has likewise been synthesized in the predicted structure but was experimentally found to be nonmagnetic[52]. TaCrP, TiMnP, TiMnAs, NbMnSi, TaMnSi, NbMnGe, TaMnGe, TiFeGe, and NbFeB have been reported either to exhibit polymorphism or to crystallize in structure types other than TiNiSi (Table 3). However, most of these reports date back several decades. Renewed synthesis efforts would therefore be valuable for reassessing their crystal structures and possible polymorphism, particularly because the calculated $E_d$ values of the TiNiSi- and ZrAlNi-type structures are similar for many of these compositions (Figs. 1 and 2). The synthesis of MoMnB and WMnB may be challenging owing to the refractory nature of boron. In addition, ScCoAs and TaCrSi have relatively small calculated magnetic moments. These considerations leave six leading TiNiSi-type FM candidates: YFeP, LaFeP, ZrMnAs, HfMnAs, TiMnGe, and ZrFeGe.

**Noncollinearity**

Collinear total-energy calculations of different magnetic configurations find that 16 compounds stabilize AM ground states. However, NbMnP has been experimentally reported to exhibit a noncollinear AFM structure[38], suggesting that noncollinear ordering may also be relevant in related compounds. We therefore investigate noncollinear configurations in six representative materials, including the neutron-diffraction-characterized NbMnP[38], TaMnP[40], and WFeB[20], the synthesized but not yet neutron-diffraction-characterized NbMnAs[41], and the newly predicted TaMnAs and ScMnP, using constrained-moment noncollinear DFT without SOC. For each magnetic phase, the structure is fully relaxed within the GGA framework. A penalty functional constrains the directions of the local moments while allowing their magnitudes and the electronic structure to relax self-consistently. Several double-Q states are generated by rotating the magnetic sublattices from the collinear AM state toward distinct FM, AFM-1, and AFM-2 configurations, and their relative stability is evaluated from the total-energy variation along these canting paths.

To examine possible noncollinear magnetic states, we considered three double-Q canting paths for each compound. Starting from the collinear AM state at $\theta = 0°$, the local moments were continuously rotated toward the FM, AFM-1, or AFM-2 state at $\theta = 90°$, with their directions constrained using a penalty functional. The relative energy per formula unit was defined as

$$\Delta E(\theta) = E(\theta) - E(0), \quad (1)$$

where $E(0)$ is the energy of the collinear AM state. The resulting energy profiles were fitted to

$$\Delta E(\theta) = 2J\sin^2\theta + 4K\sin^4\theta, \quad (2)$$

where $J$ and $K$ represent the leading quadratic and higher-order quartic contributions to the energy landscape. For Eq. (2), an intermediate stationary point satisfies

$$\sin^2\theta = -\frac{J}{4K}, \quad (3)$$

so it requires $J$ and $K$ to have opposite signs with $|K/J| > 0.25$, equivalently $K/J < -0.25$. Table 4 summarizes the fitted effective exchange parameters $J$ and $K$, together with the RMS residuals, for the three canting paths of all six compounds. The corresponding average self-consistent magnetic-moment magnitude per magnetic atom at $\theta = 45°$ is also listed in units of $\mu_B$.

**Table 4. Fitted $J$-$K$ model parameters.** Fitted effective exchange parameters $J$ and $K$ obtained using $\Delta E = J[1 - \cos(2\theta)] + K[1 - \cos(2\theta)]^2 = 2J\sin^2\theta + 4K\sin^4\theta$ for the three canting paths in the six compounds. $m(\theta = 45°)$ denotes the average magnetic moment per magnetic atom at $\theta = 45°$. The RMS residual is the root-mean-square residual of the fit.

| Compound | Path | $m(\theta = 45°)$ ($\mu_B$) | $J$ (eV) | $K$ (eV) | RMS residual (eV) |
|---|---|---|---|---|---|
| NbMnAs | AM–AFM-1 | 1.94 | 0.308 | -0.039 | 0.0012 |
| | AM–AFM-2 | 2.07 | 0.172 | -0.040 | 0.0021 |
| | AM–FM | 1.98 | 0.133 | -0.009 | 0.0025 |
| TaMnAs | AM–AFM-1 | 1.72 | 0.307 | -0.046 | 0.0018 |
| | AM–AFM-2 | 1.88 | 0.167 | -0.031 | 0.0026 |
| | AM–FM | 1.76 | 0.169 | -0.039 | 0.0037 |
| NbMnP | AM–AFM-1 | 1.37 | 0.201 | -0.036 | 0.0006 |
| | AM–AFM-2 | 1.51 | 0.103 | -0.017 | 0.0024 |
| | AM–FM | 1.38 | 0.070 | -0.019 | 0.0012 |
| ScMnP | AM–AFM-1 | 2.08 | -0.003 | 0.007 | 0.0017 |
| | AM–AFM-2 | 1.92 | 0.122 | -0.023 | 0.0011 |
| | AM–FM | 1.93 | 0.008 | 0.004 | 0.0011 |
| TaMnP | AM–AFM-1 | 1.18 | 0.180 | -0.034 | 0.0011 |
| | AM–AFM-2 | 1.32 | 0.094 | -0.014 | 0.0016 |
| | AM–FM | 0.97 | 0.038 | -0.018 | 0.0019 |
| WFeB | AM–AFM-1 | 1.03 | 0.158 | -0.030 | 0.0014 |
| | AM–AFM-2 | 1.16 | 0.105 | -0.012 | 0.0006 |
| | AM–FM | 1.16 | 0.043 | -0.002 | 0.0003 |

Figure 5(a) shows the canting-energy profiles along the AM–AFM-1 path. The DFT results are well reproduced by the fitted model. For NbMnAs, NbMnP, TaMnAs, TaMnP, and WFeB, the fitted energy profiles increase monotonically with the canting angle. ScMnP exhibits the weakest energy variation; its fitted $J$-$K$ parameters (see Table 4) formally produce a very shallow intermediate minimum, but the corresponding energy scale is smaller than the RMS fitting residual and no such minimum is resolved in the calculated DFT points. Thus, this feature does not provide robust evidence for stabilization of an intermediate noncollinear state. NbMnAs and TaMnAs exhibit the steepest energy increases, indicating the largest energetic penalty for canting, whereas NbMnP, TaMnP, and WFeB display intermediate behavior. Importantly, no calculated intermediate noncollinear configuration along this path lies below the AM energy, confirming that the collinear AM state remains the lowest-energy configuration along the AM–AFM-1 rotation.

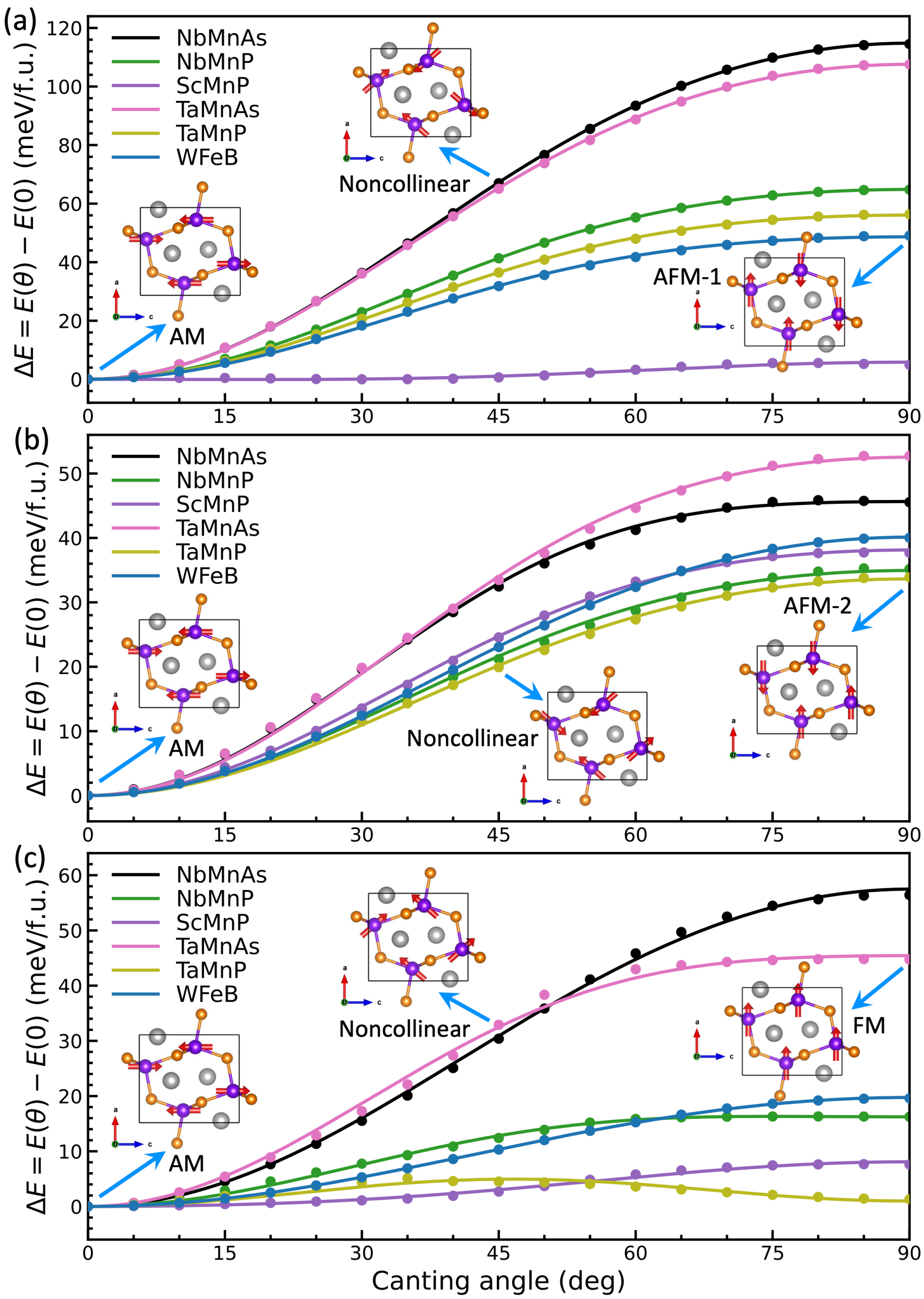

**Fig. 5. Noncollinearity.** Relative total-energy difference (Eq. (1)) as a function of the canting angle $\theta$ for NbMnAs, NbMnP, ScMnP, TaMnAs, TaMnP, and WFeB along the continuous rotation path from the collinear AM state at $\theta = 0°$ to the (a) AFM-1 state, (b) AFM-2 state, and (c) FM state, respectively, at $\theta = 90°$. The energy of the AM state is taken as the reference. The solid circles represent the constrained noncollinear DFT results, while the solid curves are fits to Eq. (2). The insets illustrate representative spin structures at $\theta = 0°$ (AM), an intermediate canting angle (noncollinear state), and $\theta = 90°$ ((a) AFM-1, (b) AFM-2, or (c) FM).

The AM–AFM-2 canting-energy profiles shown in Fig. 5(b) increase smoothly and monotonically toward the AFM-2 limit for all six compounds. No intermediate canting angle is energetically lower than the AM state. For NbMnP, the configuration at $\theta$ = 40°, which is close to the experimentally reported noncollinear structure[38], lies 18.42 meV/f.u. above the AM state. The calculated magnetic moments at this angle are nevertheless consistent with the experimental ordered moment of 1.2(1) $\mu_B$[38]. Thus, although the moment magnitude agrees with experiment, the total-energy DFT calculations favor the collinear AM state over the reported noncollinear configuration.

To investigate this discrepancy and rule out the possibility of an intermediate noncollinear stable state, we calculated the AM–AFM-2 canting-energy profile of NbMnP with experimental fixed structure using several computational treatments, including different exchange-correlation functionals, SOC, and on-site Coulomb interactions, and fitted the resulting energy profiles to the $J$-$K$ model in Eq. (2). In all cases, the energy increases monotonically from the AM state toward the AFM-2 state. The fitted $J$ and $K$ values are 0.109 and -0.009 eV (GGA+SOC), 0.060 and 0.004 eV (LDA+SOC), and 0.126 and -0.003 eV (LDA+U (U = 2 eV)), respectively. In all cases, $J > 0$ and $K/J \geq -0.083$, which is well above the critical threshold of $K/J < -0.25$ (see discussion near Eq. (3)), confirming the strictly monotonic character of the AM–AFM-2 canting-energy profile. Structural relaxation of the experimental geometry also has only a minor effect on the structure. The GGA-relaxed lattice parameters differ from the experimental values by only 1.14%, -0.40%, and 0.10% along the $a$, $b$, and $c$ axes, respectively, while the maximum internal atomic displacement is approximately 0.03 Å. These results show that the collinear AM state remains the lowest-energy configuration along the investigated AM–AFM-2 canting path in NbMnP, with no evidence for stabilization of an intermediate noncollinear state across the computational treatments considered.

Figure 5(c) presents the AM–FM canting paths. For NbMnAs, TaMnAs, ScMnP, and WFeB, the fitted energy profiles increase monotonically from the AM state toward the FM limit. For NbMnP, the energy increases from the AM state and becomes nearly saturated as it approaches the FM limit, with all calculated configurations remaining above the AM reference. TaMnP, in contrast, shows a clear nonmonotonic variation, with an initial increase followed by a decrease toward the FM state. The fitted $J$-$K$ parameters (see Table 4) for TaMnP indicate an intermediate maximum rather than a minimum, and the energy remains above the AM reference throughout the rotation. Therefore, neither the intermediate noncollinear configurations nor the FM state becomes more stable than the collinear AM state in any of the six compounds.

Taken together, the constrained noncollinear DFT calculations and fitted $J$ and $K$ values show no robust evidence for an energetically favored intermediate noncollinear state along any of the three canting paths in the six compounds. Although the associated energy scale varies across the family, with the largest values found for NbMnAs and TaMnAs and the smallest for ScMnP, the collinear AM state remains the lowest-energy configuration along all three investigated canting paths in all six compounds.

## Electronic structure

Our calculations establish a collinear AM ground state across the family, allowing the spin-polarized electronic structure to be analyzed within that order. The corresponding electronic structure of WFeB was reported previously[20]. Figure 6 presents the spin-polarized AM bands of the other three experimentally known compounds—NbMnP, TaMnP, and NbMnAs—and the six new candidates ScMnP, TaMnAs, ScMnAs, MoMnAs, MoMnSi, and WMnSi. The splitting appears along off-high-symmetry k paths outside the nodal planes or lines. The $[C_2||M_x t]$ and $[C_2||M_z t]$ nonrelativistic spin-group operations connect opposite-spin sublattices in real space; correspondingly, the momentum-space bands are related by $[C_2||M_x]$ and $[C_2||M_z]$ operations. These relations are consistent with the *d*-wave ${}^2m_x{}^1m_y{}^2m_z$ spin point group[1]. As in WFeB, every member shows a moderate AM splitting of roughly 100 meV near the Fermi level. All of these *d*-wave altermagnets are metallic, making the family suitable for spintronic applications.

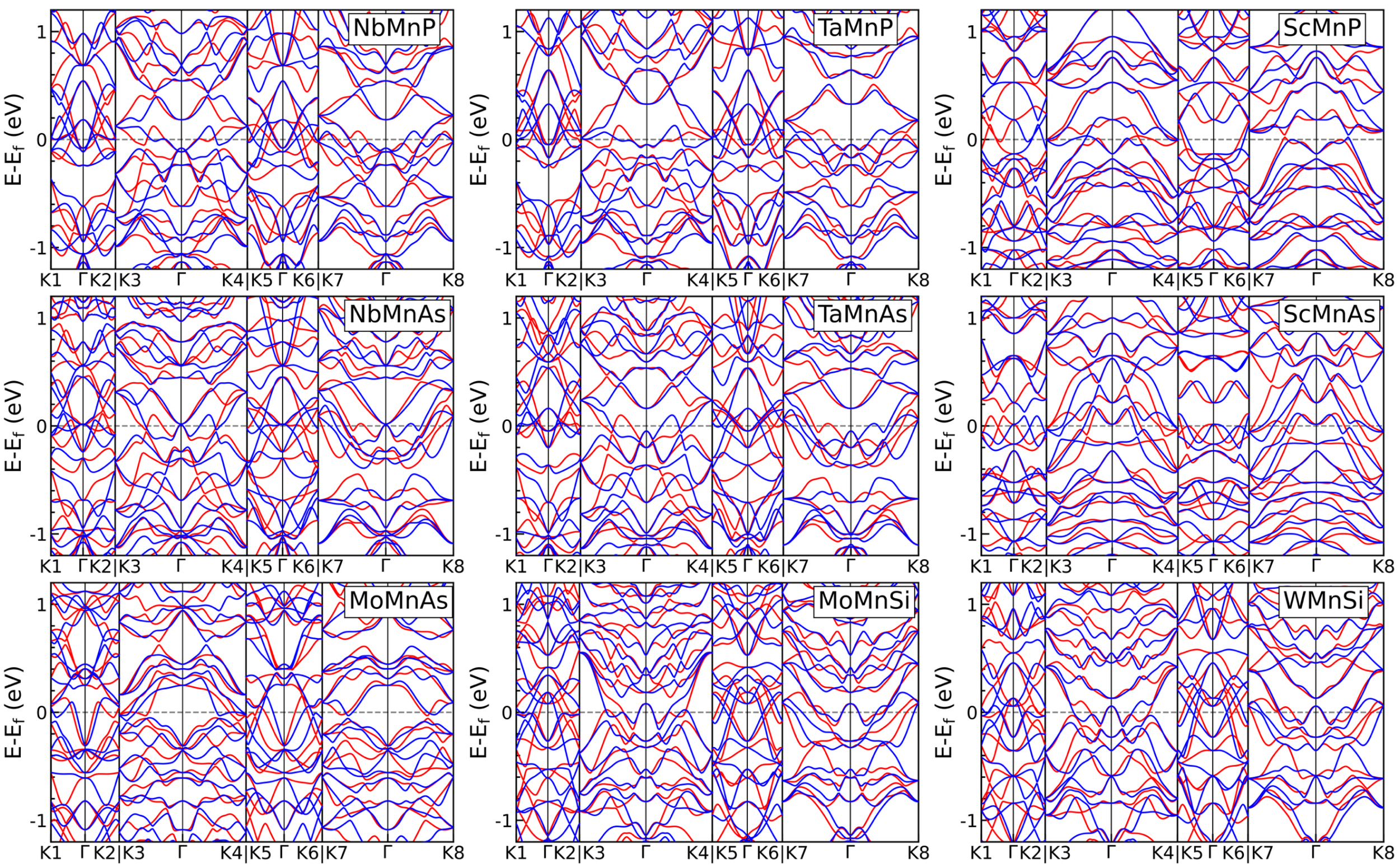

**Fig. 6. Spin-polarized band structures of the AM TiNiSi-type compounds NbMnP, TaMnP, ScMnP, NbMnAs, TaMnAs, ScMnAs, MoMnAs, MoMnSi, and WMnSi.** Blue and red curves show the bands with opposite spins. Off-high-symmetry *k* points: K1(1/4, 0, 1/2), K2(-1/4, 0, 1/2), K3(1/4, 1/2, 1/2), K4(-1/4, 1/2, 1/2), K5(1/2, 0, 1/4), K6(1/2, 0, -1/4), K7(1/2, 1/2, 1/4), and K8(1/2, 1/2, -1/4).

## Spin-splitter effect

We next examine two transport signatures relevant to spintronics—the spin-splitter effect and the AHE—in this family of metallic *d*-wave altermagnets. The spin-splitter effect and AHE arise from different parts of the conductivity response. The spin-splitter response comes from the symmetric, nonrelativistic spin-conductivity tensor component governed by the spin-point-group symmetry; the AHE instead comes from the SOC-driven antisymmetric charge-conductivity tensor component fixed by magnetic-point-group symmetry. Our previous work has evaluated the spin-splitter effect in TaMnP, NbMnAs, and WFeB[20]. The AHE has been measured for NbMnP[39], TaMnP[40], and NbMnAs[41], and has been predicted for WFeB[20]. Here, we compute both responses for the six leading new candidates: ScMnP, TaMnAs, MoMnAs, ScMnAs, MoMnSi, and WMnSi.

The TiNiSi-type nontrivial $d$-wave spin point group ${}^2m_x{}^1m_y{}^2m_z$ permits the symmetric spin-conductivity element $\sigma_{xz}^s = \sigma_{zx}^s$ in the tensor $\sigma_{\alpha\beta}^s$, whereas the charge-conductivity tensor $\sigma_{\alpha\beta}$ remains diagonal. Consequently, a charge current along $x$ or $z$ produces a transverse spin current in the $xz$ plane. This conversion is the spin-splitter effect[7]. It offers a mechanism for driving magnetization reversal in an altermagnet/ferromagnet bilayer, analogous to spin-Hall torque in a nonmagnet/ferromagnet stack[61]. Altermagnets can provide comparatively large charge-to-spin conversion without requiring strong SOC. Moreover, the spin polarization follows the Néel vector and can acquire the out-of-plane component needed for perpendicular magnetization switching[62].

We obtain the nonrelativistic spin and charge conductivities from a constant-relaxation-time Boltzmann treatment[63]. Taking a common relaxation time $\tau$ for every electronic state, the spin conductivity is

$$\sigma_{\alpha\beta}^s = \tau \sum_n \int \, v_{n\alpha} v_{n\beta} \sigma_n \frac{\partial f(E_n)}{\partial \mu} \frac{d^3k}{(2\pi)^3}. \tag{4}$$

Here, $\mathbf{v}_n(\mathbf{k})$ is the group velocity, while $\sigma_n = \pm 1$ gives the spin projection of band $n$. Removing the $\sigma_n$ factor yields the corresponding charge-conductivity formula. Because the relaxation time cancels in the spin-splitter angles $\theta_{SS} = \sigma_{zx}^s/\sigma_{xx}$ or $\sigma_{xz}^s/\sigma_{zz}$, the denominator simply identifies the charge-current direction. Because their longitudinal conductivities are nearly isotropic, the choice of charge-current direction has little qualitative influence on $\theta_{SS}$. We evaluated $\sigma_{\alpha\beta}^s/\tau$ and $\sigma_{\alpha\beta}/\tau$ with the smooth Fourier-interpolation scheme[64] available in the BoltzTraP2 package[65]. A Fermi temperature of 300 K was used to improve Brillouin-zone convergence. Repeating the calculation at lower temperatures introduced only small numerical fluctuations in the energy-dependent spin conductivity and left its Fermi-level value essentially unchanged.

Figure 7 presents the energy dependence of $\theta_{SS} = \sigma_{zx}^s/\sigma_{xx}$ for the six compounds. Near the Fermi level, $\theta_{SS}$ is approximately 0.2 in ScMnP and TaMnAs, compared with a spin Hall angle of order 0.1 for platinum, a standard material in spin-orbit-torque devices[61]. $\theta_{SS}$ is slightly below 0.1 for ScMnAs, MoMnSi, and WMnSi and is negligible only for MoMnAs. In the five cases where $\theta_{SS}$ is significant, ScMnP, TaMnAs, ScMnAs, MoMnSi, and WMnSi, $\sigma_{zx}^s$ retains the same sign over a broad energy window, suggesting tolerance to moderate doping and disorder. Thus, the TiNiSi-type metallic $d$-wave altermagnet family generally provides a useful spin-splitter response despite relatively modest AM band splitting.

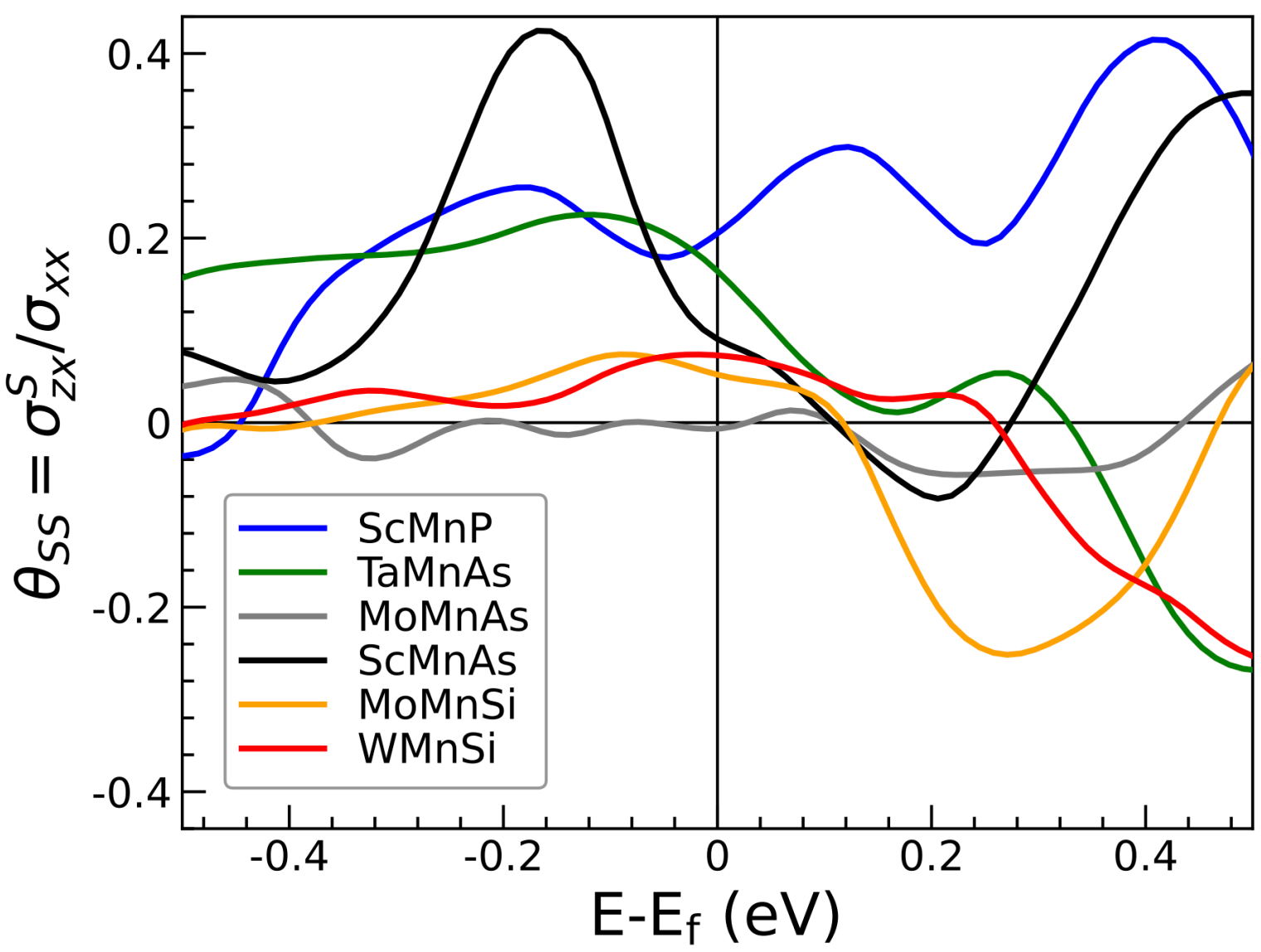

**Fig. 7. Spin-splitter angles $\boldsymbol{\theta_{SS} = \sigma^{s}_{zx}/\sigma_{xx}}$ for the AM TiNiSi-type compounds ScMnP, TaMnAs, MoMnAs, ScMnAs, MoMnSi, and WMnSi.**

**Anomalous Hall effect**

Altermagnets with $d$-wave[20] or $g$-wave[21] symmetry can also exhibit a finite anomalous Hall conductivity in the presence of SOC when the response is allowed by magnetic-point-group symmetry, providing a route to realizing the AHE in antiferromagnets. It is worth noting that altermagnets constitute the collinear class of spin-split antiferromagnets, whereas anomalous-Hall antiferromagnets are defined by their finite anomalous Hall response and may be either collinear, noncollinear, or noncoplanar. These categories therefore overlap but are not equivalent[66].

We downfolded the DFT bands onto M-$d$, T-$d$, and A-$p$ orbitals to construct maximally localized Wannier functions with the WANNIER90 code[67]. Within the $\pm 1$ eV window about the Fermi level, the resulting tight-binding Hamiltonian closely reproduces the DFT bands. We then calculated the Berry curvature from the Kubo expression[68]

$$\Omega_n^{\alpha\beta}(\mathbf{k}) \;=\; -2\,\mathrm{Im}\sum_{m\neq n}\frac{\langle u_{n\mathbf{k}}|\hat{v}_\alpha|u_{m\mathbf{k}}\rangle\langle u_{m\mathbf{k}}|\hat{v}_\beta|u_{n\mathbf{k}}\rangle}{(E_{m\mathbf{k}}-E_{n\mathbf{k}})^2}. \tag{5}$$

Here, $\hat{v}_\alpha = \frac{1}{\hbar}\partial\hat{H}/\,\partial k_\alpha$ denotes the velocity operator, and $E_{n\mathbf{k}}$ denotes the band eigenvalues. We used WannierTools[69] to integrate the occupied-state Berry curvature and obtain the intrinsic anomalous Hall conductivity

$$\sigma_{\alpha\beta}^{\mathrm{AHE}} \;=\; -\frac{e^2}{\hbar}\sum_n^{\mathrm{occ}}\int_{\mathrm{BZ}}\frac{d^3\mathbf{k}}{(2\pi)^3}\,\Omega_n^{\alpha\beta}(\mathbf{k}). \tag{6}$$

The Brillouin-zone integration employed a $100 \times 100 \times 100$ $k$-point mesh.

For the family of TiNiSi-type altermagnets, when the Néel vector $\mathbf{L}$ is oriented along [010], symmetry requires every anomalous Hall conductivity component to vanish. Orienting the Néel vector along [100] or [001] instead permits finite $\sigma^{A}_{xy}$ and $\sigma^{A}_{yz}$ components, respectively. Figure 8 plots these tensor components as functions of chemical potential. Both vary strongly with energy.

At the Fermi level, $\sigma^A_{xy}$ reaches several hundred S/cm in ScMnP, TaMnAs, MoMnAs, and WMnSi, while $\sigma^A_{yz}$ reaches comparable magnitudes in all six compounds.

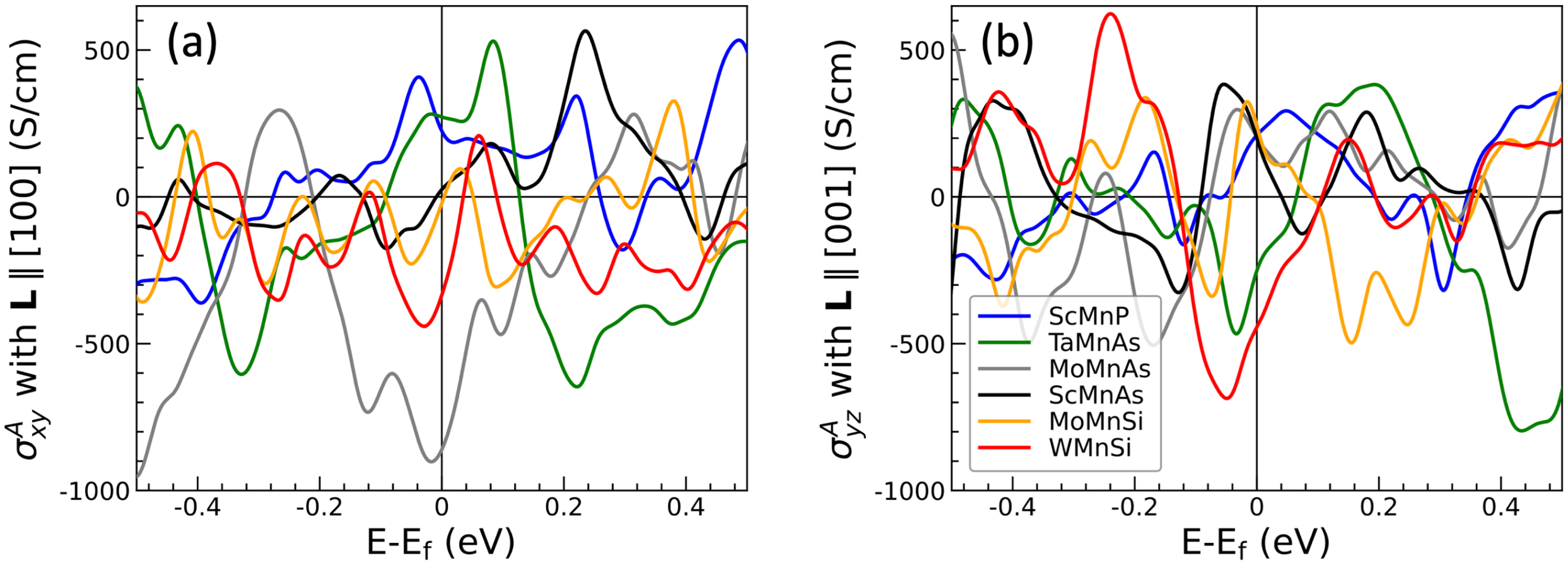

**Fig. 8. Anomalous Hall conductivities of the AM TiNiSi-type compounds ScMnP, TaMnAs, MoMnAs, ScMnAs, MoMnSi, and WMnSi.** (a) $\sigma^A_{xy}$ with **L** ∥ [100]; (b) $\sigma^A_{yz}$ with **L** ∥ [001].

## Discussion

An empirical picture for the emergence of *d*-wave altermagnetism in the TiNiSi structure is provided by a specific magnetic motif—ferromagnetically ordered zigzag chains with antiferromagnetic interchain coupling in a relatively low-symmetry crystal environment—which limits the number of rotation and mirror operations relating the opposite-spin sublattices, here represented by the ferromagnetically ordered zigzag chains (see Fig. S2 of the Supplementary Information). Consequently, spin-momentum locking divides the Brillouin zone into fewer alternating spin sectors than in higher-order *g*-wave and *i*-wave altermagnets[1]. This empirical magnetic motif may serve as a useful heuristic for high-throughput searches of materials databases[44]. Nevertheless, the final identification of *d*-wave altermagnetism requires spin-point-group analysis and band-structure verification. Moreover, representing the magnetic motif using only the magnetic atoms is a simplified description. The arrangement of the magnetic atoms is accommodated by the nonmagnetic atoms, which are essential for constraining the altermagnetic electronic structure[2,21].

The TiNiSi structure type combines this magnetic motif with broad chemical versatility, motivating a high-throughput first-principles assessment of thermodynamic stability and magnetic ground states across 280 TiNiSi-type MTA systems (M = Sc, Ti, Y, Zr, Nb, Mo, La, Hf, Ta, W; T = Cr, Mn, Fe, Co; A = N, P, As, C, Si, Ge, B). The screening identifies 121 (183) compounds with $0 \leq E_d \leq 0.05$ eV/atom ($0 \leq E_d \leq 0.2$ eV/atom). Such phases occur for A = P, As, Si, Ge, and B, but not for N or C. Among the 183 stable or metastable compounds within 0.2 eV/atom of the convex hull, 84 are nonmagnetic, 85 retain well-defined moments, and 14 show weak magnetism; six of the latter may host altermagnetic spin fluctuations. The 85 robust magnets separate into 16 *d*-wave altermagnets, 37 ferromagnets, and 32 conventional antiferromagnets. Recovering all four experimentally known altermagnets or related antiferromagnets in this family—WFeB, NbMnP, TaMnP, and NbMnAs—validates the screening strategy. Of the 12 new predictions, six stand out in terms of synthesis prospects: ScMnP, TaMnAs, ScMnAs, MoMnAs, MoMnSi, and WMnSi. Systematic noncollinearity calculations for representative established and predicted compounds, WFeB, NbMnP, TaMnP, NbMnAs, ScMnP, and TaMnAs, continue to favor the collinear altermagnetic state. Band calculations show metallicity and modest *d*-wave altermagnetic spin splitting of roughly 100 meV throughout the family. The six leading new metallic *d*-wave altermagnets exhibit finite spin-splitter angles and anomalous Hall conductivities. Remarkably, ScMnP and TaMnAs reach spin-splitter angles of about 0.2—roughly twice the spin Hall angle of platinum—despite their modest altermagnetic band splitting. These results enlarge the limited pool of metallic *d*-wave altermagnets available for efficient charge-to-spin conversion through altermagnetic symmetry and provide targets for experimental study.

## Methods

### First-principles calculations

The high-throughput DFT calculations employed VASP[70] with the projector-augmented-wave (PAW) method[71] and the generalized-gradient approximation (GGA)[72]. Brillouin-zone integrations used a Γ-centered $2\pi \times 0.033$ Å$^{-1}$ k-point grid, and the plane-wave kinetic-energy cutoff was 600 eV. Electronic self-consistency was converged to $10^{-5}$ eV, while ionic relaxation was continued until the residual forces were below 0.01 eV Å$^{-1}$. For the identified altermagnets, we evaluated the band structures, AHE, and spin-splitter response with the local density approximation (LDA) using the GGA-relaxed geometries. A Γ-centered $9 \times 16 \times 7$ k-point grid was used for the band calculations that provided the input for the AHE and spin-splitter computations. Magnetic exchange parameters were obtained with the TB2J package[73] from localized orbitals generated with OpenMX[74].

## Acknowledgments

This work was supported by the U.S. Department of Energy (DOE) Established Program to Stimulate Competitive Research (EPSCoR) Grant No. DE-SC0024284. Computations were performed at the High Performance Computing facility at Iowa State University and the Holland Computing Center at the University of Nebraska.

# Supplementary Information

## Mapping metallic *d*-wave altermagnetism across the TiNiSi structural family

Zhen Zhang[1,*], Subhadip Pradhan[2], Kirill D. Belashchenko[2], Vladimir Antropov[1,3,*]

[1]*Department of Physics and Astronomy, Iowa State University, Ames, IA 50011, USA*

[2]*Department of Physics and Astronomy and Nebraska Center for Materials and Nanoscience, University of Nebraska-Lincoln, Lincoln, Nebraska 68588, USA*

[3]*Ames National Laboratory, U.S. Department of Energy, Ames, IA 50011, USA*

[*]Corresponding authors: Zhen Zhang zhenz1@iastate.edu, Vladimir Antropov antropov@iastate.edu

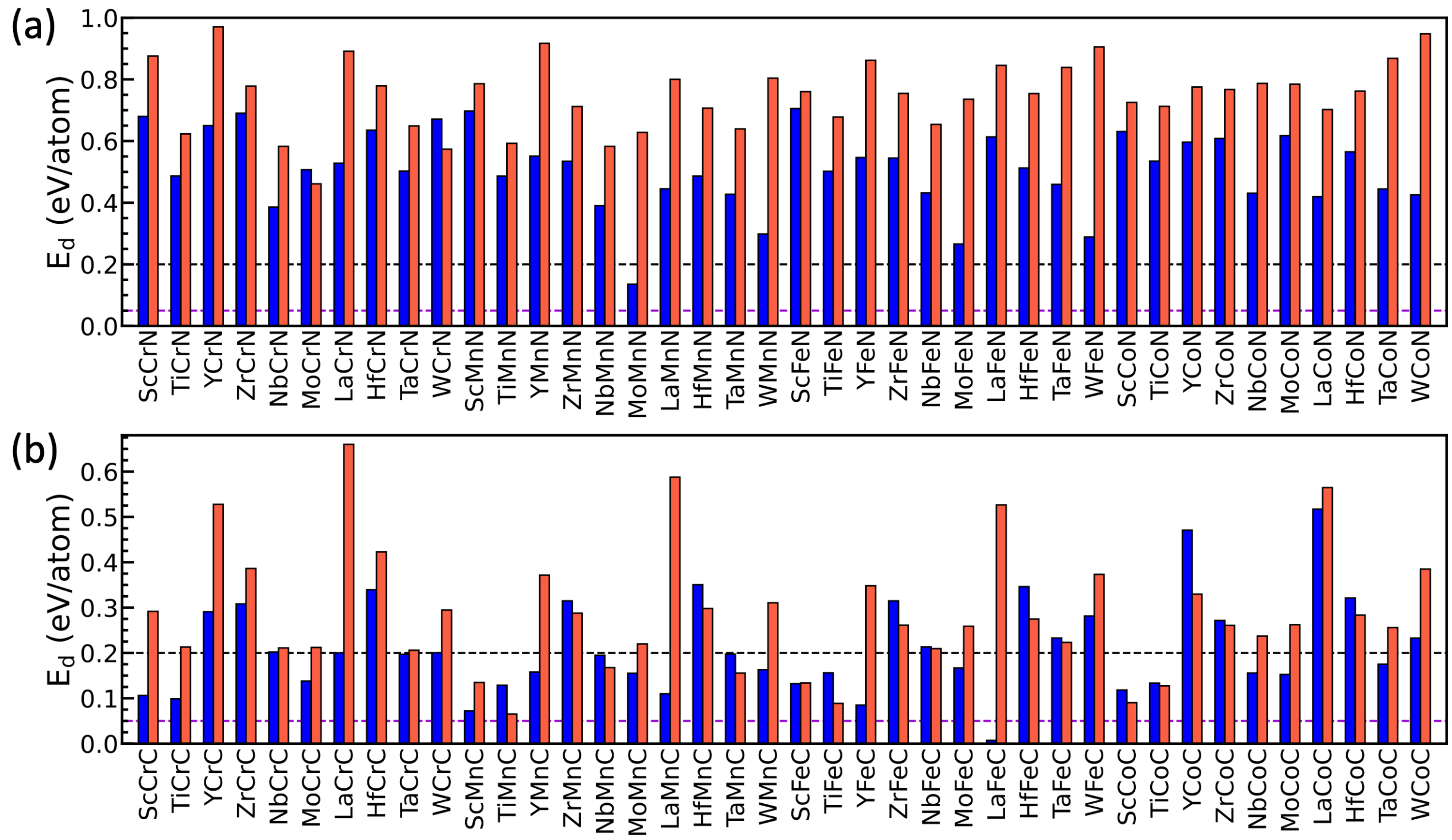

Fig. S1. Convex-hull distances for MTA compounds with A = N and C. Hull distances for (a) nitrides and (b) carbides. Blue and red poles denote TiNiSi- and ZrAlNi-type phases, respectively. The criteria $E_d \leq 0.2$ eV/atom and $E_d \leq 0.05$ eV/atom are marked by the black and purple horizontal dashed lines, respectively.

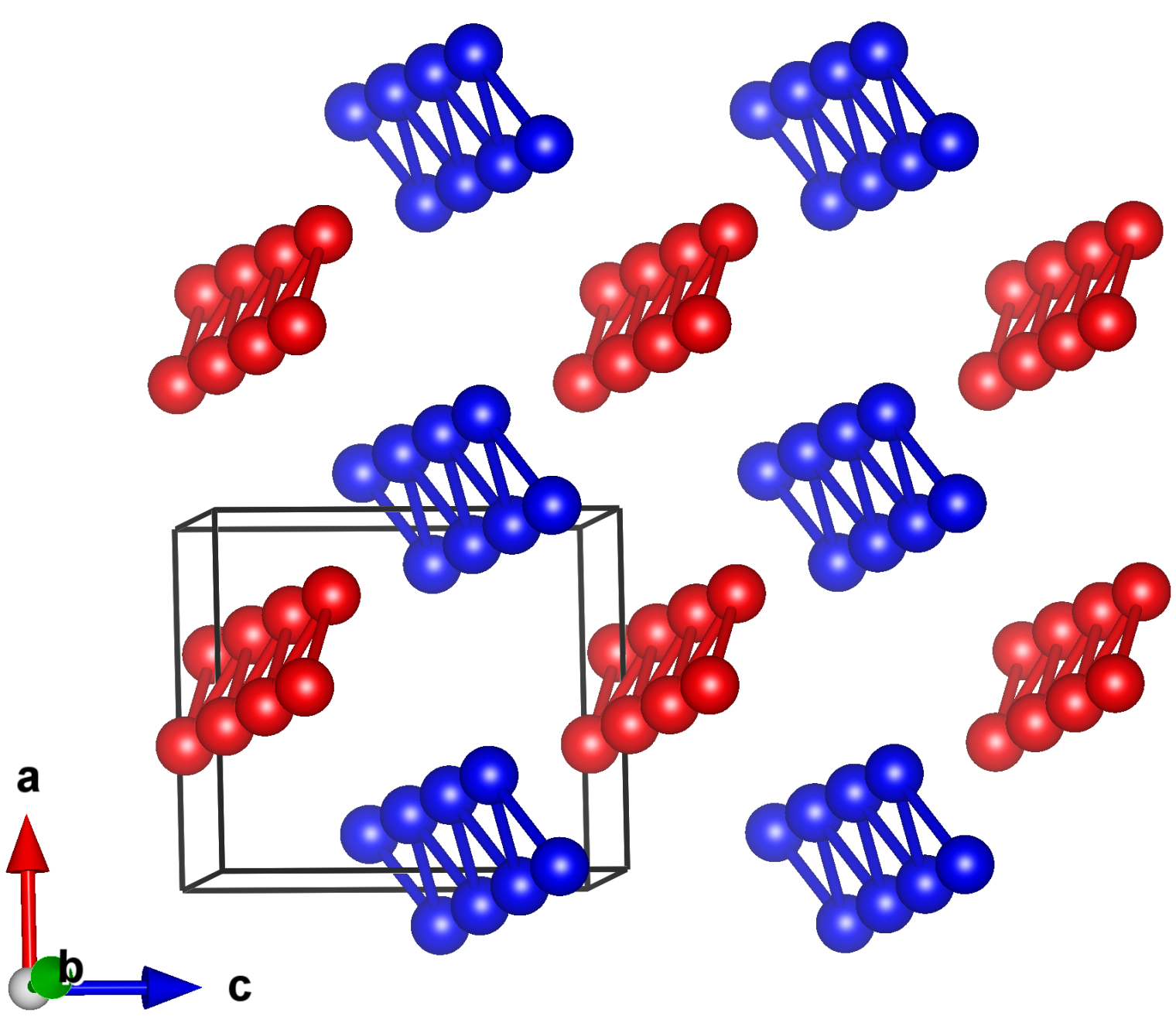

Fig. S2. Magnetic structure of TiNiSi-type *d*-wave altermagnet. Blue and red spheres represent magnetic atoms on the Ni site with opposite spins. Atoms on the Ti and Si sites are not shown.